\documentclass[11pt]{article}

\usepackage{graphicx}
\usepackage{cite}
\usepackage{mathpple}

\usepackage{geometry}
\begin{document}

{\small\noindent
\textbf{Preprint of:}\\
T. A. Nieminen, H. Rubinsztein-Dunlop and N. R. Heckenberg\\
``Multipole expansion of strongly focussed laser beams''\\
\textit{Journal of Quantitative Spectroscopy and Radiative Transfer}
\textbf{79--80}, 1005--1017 (2003)
}

\vspace{6mm}

\hrulefill

\vspace{9mm}

\begin{center}

\LARGE
\textbf{Multipole expansion of strongly focussed laser beams}

\vspace{3mm}

\large
T. A. Nieminen, H. Rubinsztein-Dunlop, and N. R. Heckenberg

\vspace{3mm}

\normalsize
\textit{Centre for Biophotonics and Laser Science, Department of Physics,\\
The University of Queensland, Brisbane QLD 4072, Australia}

\texttt{timo@physics.uq.edu.au}

\end{center}

\begin{abstract}

Multipole expansion of an incident radiation field---that is, representation of
the fields as sums of vector spherical wavefunctions---is essential for
theoretical light scattering methods such as the \textit{T}-matrix method and
generalised Lorenz-Mie theory (GLMT). In general, it is theoretically
straightforward to find a vector spherical wavefunction representation of an
arbitrary radiation field. For example, a simple formula results in the useful
case of an incident plane wave. Laser beams present some difficulties. These
problems are not a result of any deficiency in the basic process of spherical
wavefunction expansion, but are due to the fact that laser beams, in their
standard representations, are not radiation fields, but only approximations of
radiation fields. This results from the standard laser beam representations
being solutions to the paraxial scalar wave equation. We present an efficient
method for determining the multipole representation of an arbitrary focussed
beam.

\vspace{3mm}


\end{abstract}

\section{Introduction}

Multipole expansion of an incident radiation field in terms of
vector spherical wavefunctions (VSWFs) is required for theoretical scattering
methods such as the \textit{T}-matrix
method~\cite{waterman1971,tsang1985,mishchenko1991} and generalised Lorenz-Mie
theory (GLMT)~\cite{lock1995}. Since the VSWFs are a complete orthogonal basis
for solutions of the vector Helmholtz equation,
\begin{equation}
\nabla^2 \mathbf{E} + k^2 \mathbf{E} = 0,
\label{helmholtz}
\end{equation}
it is theoretically
straightforward to find the multipole expansion for any monochromatic radiation
field using the orthogonal eigenfunction transform~\cite{bohren,ren1998}, also
known as the generalised Fourier transform. For example, the usual formula for
multipole expansion of a plane wave can be derived in this way.

While the case of plane wave illumination is useful for a wide range of
scattering problems, many applications of scattering involve laser
beams. In particular, laser trapping~\cite{ashkin} requires strongly focussed
beams. As scattering calculations allow the optical forces and torques within
the optical trap to be determined~\cite{nieminen2001jqsrt,polaert1998oc}, and
the \textit{T}-matrix method is particularly useful for the repeated
calculations typically required, multipole expansion of laser beams, including
strongly focussed beams, is extremely useful.

Unfortunately, laser beams present some serious theoretical difficulties.
These problems are not a result of any deficiency in
the basic process of multipole expansion of
radiation fields, but are due to the fact that standard
representations of laser beams are not radiation fields---that is, their
standard mathematical forms are not solutions of the vector
Helmholtz equation, but are solutions of the paraxial scalar wave equation
(higher order corrections can be used to improve the accuracy as reality becomes
less paraxial~\cite{davis1979}). Such
pseudo-fields are only approximate solutions of the vector wave equation. The
deviation from correctness increases as the beam is more strongly
focussed~\cite{sheppard1998,sheppard2001,ulanowski2000}.

While, in principle, any radiation field can be expanded in terms of
multipoles, multipole expansions do not exist for approximate pseudo-fields
that do not satisfy the vector wave equation. Since the standard paraxial
representations of laser beams are so widely used, multipole expansions
that correspond to the standard beams in a meaningful way are highly
desirable. For this to be possible, some method must be used
to ``approximate'' the standard laser beam with a real radiation field.
We can note that there exists a significant and useful body of work on
multipole expansion of
beams~\cite{doicu1997ao,gouesbet1995,gouesbet1996b,%
lock1996b,lock1996c,polaert1998ao,han2001}. However, while satisfactorily
efficient and accurate methods exist for weakly focussed beams, when the
deviations from paraxiality are small, strongly focussed beams, for example, as
required for laser trapping, remain problematic.

The traditional method used in scattering calculations was to assume that the
actual incident field was equal to the paraxial field on the surface of the
scatterer. This has the unfortunate drawback that the multipole expansion
coefficients depend on the size, shape, location, and orientation of the
scattering particle. 
As a result, an artificial dependence on particle position and size
is introduced into scattering calculations and calculations of optical forces,
since the differing multipole expansions correspond to different beams. 
This is obviously undesirable.
While matching the fields on the surface of the scattering particle
is not an adequate solution to the problem, the
basic concept---matching the fields on a surface---is sound. However, a
scatterer-independent surface must be used. Two natural choices present
themselves: the focal plane (for beams with a well-defined focal plane), and a
spherical surface in the far field.

Integral methods can be used over these surfaces (direct application of the
orthogonal eigenfunction transform). Here, however, we use a point-matching
method since, firstly, the method tolerates a much coarser grid of points,
thereby increasing computational efficiency, and secondly, increased robustness
can be gained by using extra points to give an overdetermined system of
equations which can then be solved in a least-squares sense. We use both focal
plane matching and far field matching, and evaluate their respective merits.

\section{Point-matching method}

All monochromatic radiation fields satisfy the vector Helmholtz
equation~(\ref{helmholtz}) in source-free regions. The vector spherical
wavefunctions (sometimes called vector spherical harmonics) are a complete set
of orthogonal solutions to this equation. The VSWFs are
\begin{eqnarray}
\mathbf{M}_{nm}^{(1,2)}(k\mathrm{r}) & = & N_n h_n^{(1,2)}(kr)
     \mathbf{C}_{nm}(\theta,\phi) \\
\mathbf{N}_{nm}^{(1,2)}(k\mathrm{r}) & = & \frac{h_n^{(1,2)}(kr)}{krN_n}
\mathbf{P}_{nm}(\theta,\phi) + \nonumber \\
& & N_n \left( h_{n-1}^{(1,2)}(kr) -
\frac{n h_n^{(1,2)}(kr)}{kr} \right) \mathbf{B}_{nm}(\theta,\phi)
\label{vswfn}
\end{eqnarray}
where $h_n^{(1,2)}(kr)$ are spherical Hankel functions of the first and
second kind, $N_n = 1/\sqrt{n(n+1)}$ are normalisation constants, and
$\mathbf{B}_{nm}(\theta,\phi)$, $\mathbf{C}_{nm}(\theta,\phi)$, and
$\mathbf{P}_{nm}(\theta,\phi)$ are the vector spherical harmonics:
\begin{eqnarray}
\mathbf{B}_{nm}(\theta,\phi) & = & \mathbf{r} \nabla
Y_n^m(\theta,\phi) = \nabla \times \mathbf{C}_{nm}(\theta,\phi) \nonumber \\
& = & \hat\theta
\frac{\partial}{\partial\theta} Y_n^m(\theta,\phi) + \hat\phi
\frac{\mathrm{i}m}{\sin\theta} Y_n^m(\theta,\phi),\\
\mathbf{C}_{nm}(\theta,\phi) & = & \nabla \times \left( \mathbf{r}
Y_n^m(\theta,\phi) \right) \nonumber \\
& = & \hat\theta\frac{\mathrm{i}m}{\sin\theta}
Y_n^m(\theta,\phi) - \hat\phi \frac{\partial}{\partial\theta}
Y_n^m(\theta,\phi),\\
\mathbf{P}_{nm}(\theta,\phi) & = & \hat{\mathbf{r}} Y_n^m(\theta,\phi),
\end{eqnarray}
where $Y_n^m(\theta,\phi)$ are normalised scalar spherical harmonics. The usual
polar spherical coordinates are used, where $\theta$ is the co-latitude measured
from the $+z$ axis, and $\phi$ is the azimuth, measured from the $+x$ axis
towards the $+y$ axis. The $h_{n-1}^{(1,2)}(kr) - n h_n^{(1,2)}(kr)/kr$ term in
(\ref{vswfn}) results from the identity $(\mathrm{d}/\mathrm{d}kr)
krh_n^{(1,2)}(kr) = kr h_{n-1}^{(1,2)}(kr) - n h_n^{(1,2)}(kr)$.

$\mathbf{M}_{nm}^{(1)}$ and $\mathbf{N}_{nm}^{(1)}$ are outward-propagating TE
and TM multipole fields, while $\mathbf{M}_{nm}^{(2)}$ and
$\mathbf{N}_{nm}^{(2)}$ are the corresponding inward-propagating multipole
fields. Since these wavefunctions are purely incoming and purely outgoing, each
has a singularity at the origin. Since the incident field has equal incoming and
outgoing parts, and must be singularity free in any source-free region, it is
useful to define the singularity-free regular vector spherical wavefunctions:
\begin{eqnarray}
\mathbf{RgM}_{nm}(k\mathrm{r}) & = & {\textstyle\frac{1}{2}} [
\mathbf{M}_{nm}^{(1)}(k\mathrm{r}) + \mathbf{M}_{nm}^{(2)}(k\mathrm{r}) ], \\
\mathbf{RgN}_{nm}(k\mathrm{r}) & = & {\textstyle\frac{1}{2}} [
\mathbf{N}_{nm}^{(1)}(k\mathrm{r}) + \mathbf{M}_{nm}^{(2)}(k\mathrm{r}) ].
\end{eqnarray}
Since the spherical Bessel functions $j_n(kr) = \frac{1}{2}(h_n^{(1)}(kr) +
h_n^{(2)}(kr))$, the regular VSWFs are identical to the incoming and outgoing
VSWFs except for the replacement of the spherical Hankel functions by spherical
Bessel functions.

Since the VSWFs are a complete set of solutions to the vector Helmholtz
equation~(\ref{helmholtz}), every solution can be written as a linear
combination of VSWFs. Thus, the incident field can be written in terms of
\emph{expansion coefficients} (also called \emph{beam shape coefficients})
$a^{(3)}_{nm}$ and $b^{(3)}_{nm}$ as
\begin{equation}
\mathbf{E}_\mathrm{inc}(\mathrm{r}) = \sum_{n=1}^\infty \sum_{m = -n}^n
a^{(3)}_{nm} \mathbf{RgM}_{nm}(k\mathrm{r}) +
b^{(3)}_{nm} \mathbf{RgN}_{nm}(k\mathrm{r}).
\label{regular_expansion}
\end{equation}
Alternatively, since the outgoing field is purely a consequence of the
incoming field (with or without a scatterer present), all the necessary
information can be conveyed by writing the incident field expansion
in terms of incoming wavefunctions as
\begin{equation}
\mathbf{E}_\mathrm{inc}(\mathrm{r}) = \sum_{n=1}^\infty \sum_{m = -n}^n
a^{(2)}_{nm} \mathbf{M}_{nm}^{(2)}(k\mathrm{r}) +
b^{(2)}_{nm} \mathbf{N}_{nm}^{(2)}(k\mathrm{r}).
\label{outgoing_expansion}
\end{equation}
The two sets of expansion coefficients are related, since
$a^{(3)}_{nm} = 2a^{(2)}_{nm}$ and $b^{(3)}_{nm} = 2b^{(2)}_{nm}$ (the
scattered/outgoing field expansion coefficients will differ). In practice, the
multipole expansion will be terminated at some $n = N_{\mathrm{max}}$. For
the case of multipole fields produced by an antenna that is contained
within a radius $a$, $N_{\mathrm{max}} = ka$ is usually adequate, but
$N_{\mathrm{max}} = ka + 3 \sqrt[3]{ka}$ is advisable if higher accuracy is
needed~\cite{brock}. This can also be used as a guide for choosing
$N_{\mathrm{max}}$ for beams---if the beam waist is contained in a radius $a$,
this can be used to choose $N_{\mathrm{max}}$. Since the multipole field
necessarily deviates from the paraxial field, the required accuracy is open to
question. It appears that $a = w_0$ generally gives adequate results.

For focal plane matching of the fields, the regular wavefunctions are used,
while for far field matching, the incoming wavefunctions are used. The use of
only the incoming portion of the field for matching the fields in the far field
means that no knowledge of the outgoing portion of the fields is required.

If the field expansion is truncated at $n = N_{\mathrm{max}}$,
equation~(\ref{regular_expansion}) or (\ref{outgoing_expansion}) contain
$2(N_{\mathrm{max}}^2 + 2N_{\mathrm{max}})$ unknown variables---the
expansion coefficients $a_{nm}$ and $b_{nm}$. They can be found from a set of
$2(N_{\mathrm{max}}^2 + 2N_{\mathrm{max}})$ equations. If we know the field at
enough points in space, either by using the standard paraxial form of the beam,
or from measurements, a sufficiently large set of equations can be generated.
Since the electric field has three vector components at each point in space,
three equations result from each point. For robustness, extra points can be used
to give an over-determined linear system which can then be solved in a
least-squares sense. This is the basic point-matching procedure. The point
spacing along any spherical coordinate must be sufficient to prevent aliasing of
the VSWFs with respect to that coordinate.

The linear system can be written as a matrix of size $O(N_{\mathrm{max}}^2)$, so
solution of the system will scale as $O(N_{\mathrm{max}}^6)$. While this is
clearly less desirable than integral methods which scale as
$O(N_{\mathrm{max}}^4)$ for surface methods and $O(N_{\mathrm{max}}^5)$ for
volume methods, integral methods require a much more closely spaced grid of
points. Since we are mainly interested in strongly focussed beams, when
$N_{\mathrm{max}}$ is small, the gain in speed through the use of a coarse grid
results in improved performance. We can note that the most efficient solution to
very wide beams is simply to approximate them as plane waves
(an $O(N_{\mathrm{max}})$ solution), which will be quite adequate if the
scattering particle is small compared to the beam width, and located near the
beam axis. Optimisations for axisymmetric beams, giving $O(N_{\mathrm{max}}^3)$
performance, are described later.

\section{Focal plane matching}

Point-matching in the focal plane offers a number of advantages.
Firstly, the paraxial beam has a simple phase structure in the focal plane.
Secondly, the focal plane is likely to be the region in which scattering or
trapping takes place. Thirdly, the irradiance distribution in the focal plane
may be known experimentally. Although an accurate measurement of the focal plane
irradiance distribution is usually quite difficult, due to the finite resolution
of typical detectors, the structure of the beam, and the approximate focal spot
size, can be determined. The major disadvantage is that the greatest differences
between the paraxial beam and the real beam can be expected to occur in the
focal plane. For example, a paraxial beam has zero longitudinal components of
the electric field, but, from Maxwell's equations, the longitudinal component
must be non-zero.

For mathematical convenience, we choose a coordinate system so that the beam
axis coincides with the $z$ axis, with the beam propagating in the $+z$
direction, with the focal plane coincident with the $xy$ plane (the $\theta =
\pi/2$ plane). Next, we can generate a grid of points, and calculate the
electric field components of the paraxial beam at these points.

A practical difficulty immediately presents itself, however, since the
spherical harmonics $Y_n^m(\theta,\phi)$ with odd $n + m$ are odd with
respect to $\theta$, and will be zero for $\theta = \pi/2$---the focal plane.
This is to be expected physically, since the same electric field amplitude in
the focal plane can correspond to a beam propagating in the $+z$ direction, the
$-z$ direction, or a standing wave created by the combination of these. A
mathematically rigorous way of dealing with this is to match the derivatives of
the electric field amplitude, as well as the field amplitude itself. However,
this isn't necessary, since we have already chosen a direction of propagation as
one of the initial assumptions. Therefore, it is sufficient,  and
computationally efficient, to use only those VSWFs that will be non-zero to
determine the corresponding ``missing'' expansion coefficients. This will give
half of the required expansion coefficients. Then,  from the direction of
propagation, $a_{nm} = - b_{nm}$ or $b_{nm} = - a_{nm}$ gives the ``missing''
coefficients. If the beam propagates in the $-z$ direction, $a_{nm} = b_{nm}$ or
$b_{nm} = a_{nm}$ are used. If the beam was a pure standing wave combination of
counter-propagating beams, these expansion coefficients would be zero.

In principle, this procedure can be carried out for an arbitrary beam. However,
determining the vector components of beams where the Poynting vector is not
parallel to the $z$ axis in the focal plane will be problematic, since there
will be no unique choice of electric field direction since the paraxial beam is
also a scalar beam. Therefore, we will restrict ourselves to beams with
$z$-directed focal plane Poynting vectors only. This eliminates, for example,
Laguerre-Gaussian beams LG$_{pl}$ with $l \neq 0$.

Often, the beam of interest will have an axisymmetric irradiance distribution.
Any beam that can be written as a sum of $l = 0$ Laguerre-Gaussian modes
LG$_{p0}$ will satisfy this criterion. The most significant such beam is the
TEM$_{00}$ Gaussian beam (which is also the LG$_{00}$ beam). Beams with
these properties have only $m = \pm 1$ multipole
components~\cite{gouesbet1995,gouesbet1996b}, which follows simply from the
$\exp(\mathrm{i}m\phi)$ dependence of the VSWFs. This allows a significant
computational optimisation, since only the $m = \pm 1$ VSWFs need to be included
in the point-matching procedure. This results in $O(N_{\mathrm{max}}^3)$
performance.

The point-matching procedure described above was implemented using
MATLAB~\cite{matlab}. Since the main computational step is the solution of an
overdetermined linear system, a linear algebra oriented language/system such as
MATLAB was natural choice. Additionally, existing MATLAB routines for the
calculation of Bessel, Hankel, and associated Legendre functions allow simple
calculation of VSWFs. Since the VSWFs only need to be calculated for $\theta =
\pi/2$, a special purpose fast VSWF calculation routine could be used.

\subsection{TEM$_{00}$ beam}

The scalar amplitude $U$ of a TEM$_{00}$ Gaussian beam is given in cylindrical
coordinates by~\cite{siegman}

\begin{equation}
U = U_0 \frac{q_0}{w_0+q_z} \exp \left[ \mathrm{i} k
\left( z + \frac{r^2}{2 q_z}  \right) \right]
\label{paraxial_gaussian}
\end{equation}

where $k$ is the wavenumber, $w_0$ is the waist radius, $U_0$ is the central
amplitude, $q_z = q_0 + z$ is the complex radius of curvature, and $q_0 = -
\mathrm{i} z_R$ where $z_R = k w_0^2 / 2$ is the Rayleigh range. In the focal
plane, $z = 0$ and the above expression can be slightly simplified. For a beam
linearly polarised in the $x$ direction, $E_x = U$ and $E_y = 0$, while for
circularly polarised beams, $E_x = U$ and $E_y = \pm \mathrm{i} U$. $E_z = 0$ in
all cases.

\begin{figure}[htb]
\centerline{\includegraphics[width=\columnwidth]{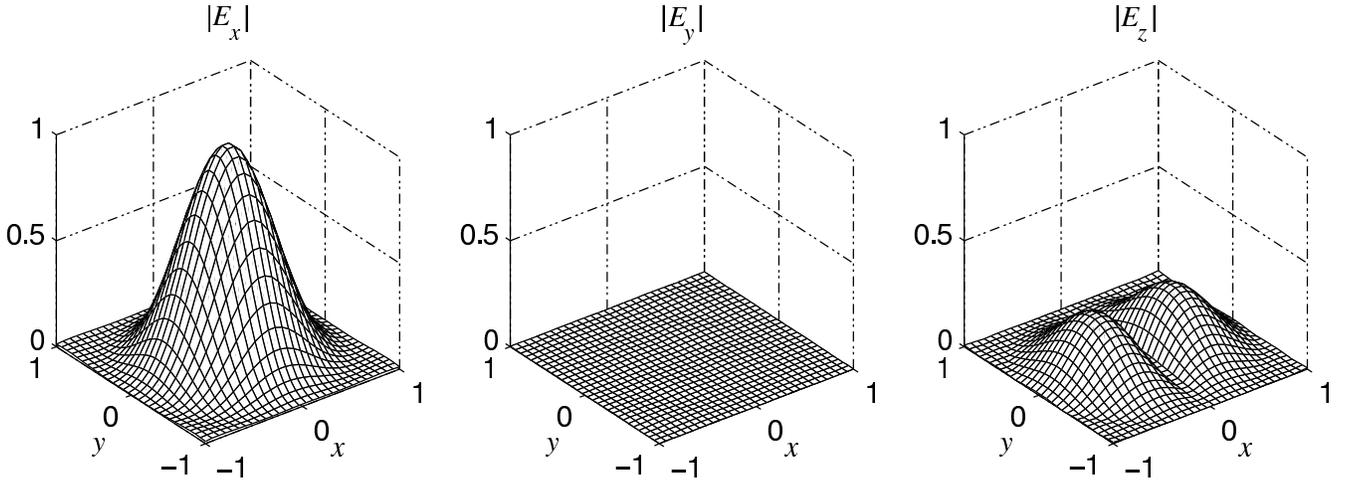}}
\caption{$x$,  $y$, and $z$ components of the electric field of a multipole
beam focal-plane matched with a TEM$_{00}$ Gaussian beam of waist radius
$w_0 = 0.5\lambda$. The fields are normalised to the central value of
$|E_x|$. All distances are in units of the wavelength.}
\label{focal_components}
\end{figure}

The general structure of a real beam point-matched with a TEM$_{00}$
Gaussian beam of waist radius $w_0 = 0.5\lambda$ is shown in
figure~\ref{focal_components}. The beam is very similar to those given by
Gouesbet's localised
approximation~\cite{gouesbet1995,gouesbet1996b,polaert1998ao},  although the
(very small) $y$ component is somewhat different. The robustness of the
algorithm is shown by the generation of the correct $z$ components for the final
beam, even though the original paraxial beam was unphysically assumed to have
$E_z = 0$.

\begin{figure}[htb]
\centerline{\includegraphics[width=\columnwidth]{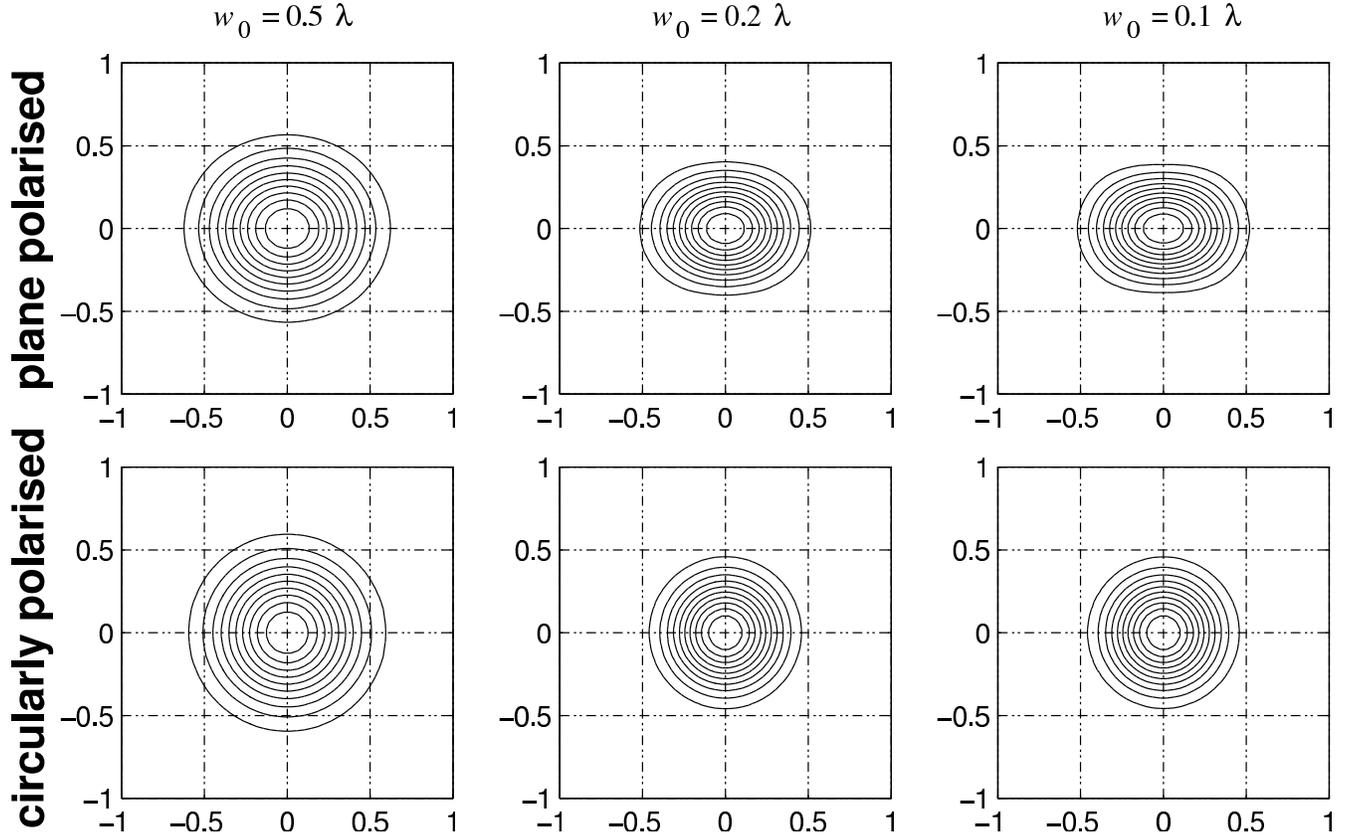}}
\caption{Irradiance contours in the focal plane of multipole beams focal-plane
matched with TEM$_{00}$ beams. The top row shows plane-polarised beams, with the
plane of polarisation in the horizontal plane, and the bottom row shows
circularly-polarised beams. The beam waists vary from left to right as
$w_0 = 0.5\lambda$, $w_0 = 0.2\lambda$, and $w_0 = 0.1\lambda$. The two main
features to note are the increasing axial asymmetry of the plane-polarised beam
as it is more strongly focussed, and the approach of the beam to a minimum focal
spot radius. All distances are in units of the wavelength.}
\label{focal_6plot}
\end{figure}

The behaviour of the beam as it is more tightly focussed is shown in
figure~\ref{focal_6plot}. The increasing axial asymmetry of
plane-polarised beams as they are more strongly focussed can be clearly seen.
The other feature to note is that focussed beams have a minimum waist
size---the more strongly
focussed beams shown here closely approach this minimum size,
the well-known diffraction limited focal spot~\cite{siegman,born,sales1998}.
The lack of any such limit in the paraxial beam is blatantly
unphysical, and we
cannot expect a beam matched to a $w_0 = 0.1\lambda$ paraxial beam to have such
an unreasonably small waist radius.

The $N_{\mathrm{max}}$ for these calculations was determined by assuming that an
effective radius of $3w_0$ contains the entire beam in the focal plane. This
gives $N_{\mathrm{max}} = 16$, 9, and 6 for the three beam widths of
$w_0 = 0.5\lambda$, $w_0 = 0.2\lambda$, and $w_0 = 0.1\lambda$. All VSWFs were
included in the solution (that is, the axisymmetric beam optimisation was not
used), so the number of unknown variables for each case was 576, 198, and 96.
The fields were matched at a number of points equal to the number of
variables, distributed on a polar grid. This is an overdetermined
system, with 50\% more points than required. The times required for the solution
of the linear system on a 1.5\,GHz PC were 11.2\,s, 0.50\,s, and 0.046\,s,
respectively. If the solution is restricted to the $m = \pm 1$ VSWFs, the
number of variables is reduced to $4N_{\mathrm{max}}$, and the times required
for the three cases shown are reduced to 0.006\,s, 0.002\,s, and 0.001\,s.

\subsection{Bi-Gaussian beam}

We consider a simple non-axisymmetric beam, constructed by replacing $r^2$ in
the paraxial Gaussian beam formula~(\ref{paraxial_gaussian}) by
$(ax)^2 + (by)^2$, giving a bi-Gaussian irradiance distribution with an
elliptical focal spot. The expansion coefficients are no longer restricted to
$m = \pm 1$.

\begin{figure}[htb]
\centerline{\includegraphics[width=\columnwidth]{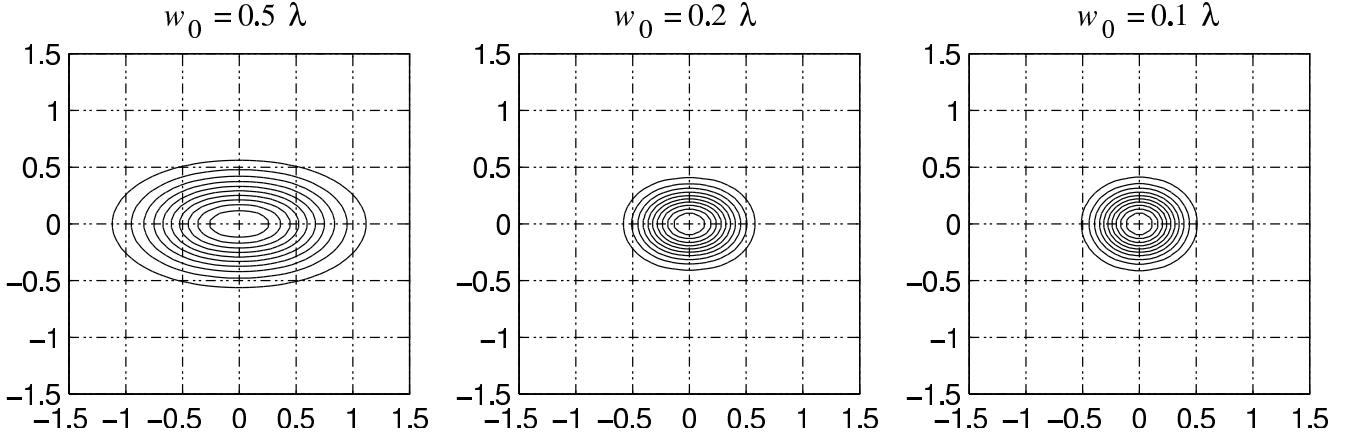}}
\caption{Irradiance contours in the focal plane of multipole beams focal-plane
matched with circularly-polarised bi-Gaussian beams. The beam ellipticity
factors are $a = 0.5$ and $b = 1$ for all cases. The beam waists vary from left
to right as $w_0 = 0.5\lambda$, $w_0 = 0.2\lambda$, and $w_0 = 0.1\lambda$.
The effective beam waist in the $x$ direction is $2w_0$. All distances
are in units of the wavelength.}
\label{focal_bigauss}
\end{figure}

The irradiance distributions of point-matched bi-Gaussian beams are shown in
figure~\ref{focal_bigauss}. The effect of the minimum spot size is evident, as
it first restricts the focussing in the $y$ direction, and then the $x$
direction,  as the beam is more strongly focussed. The odd $m$ multipole
components are non-zero in these beams. The increased width of the beam results
in longer calculations using the $3w_0$ cutoff radius. A smaller cutoff radius
equal to $w_0$ produces very similar results, and may be more suitable for
practical use. Since the non-zero expansion coefficients can be predicted from
the symmetry of the beam, it would also be possible to speed up the calculations
by only including those coefficients.

\section{Far field matching}

In the far field, the beam becomes an inhomogeneous spherical wave. The radial
component of the electric and magnetic fields vanishes. The large radius limits
for the VSWFs~\cite{mishchenko1991}:
\begin{eqnarray}
\mathbf{M}_{nm}^{(1,2)}(k\mathrm{r})|_{kr \gg n^2} & = &
\frac{N_n}{kr} (\mp \mathrm{i})^{n+1}
\exp({\pm \mathrm{i}kr}) \mathbf{C}_{nm}(\theta,\phi) \\
\mathbf{N}_{nm}^{(1,2)}(k\mathrm{r})|_{kr \gg n^2} & = &
\frac{N_n}{kr} (\mp \mathrm{i})^n
\exp({\pm \mathrm{i}kr}) \mathbf{B}_{nm}(\theta,\phi)
\end{eqnarray}
can be usefully employed. The far field is best written in spherical
polar coordinates, since this gives at most two non-zero components.
In cartesian coordinates,  all three components will generally be non-zero,
although only two will be independent. Since only two vector
components of the field
can be used in the point matching procedure,
the minimum number of points at which to match the fields
that is needed is higher.

The non-zero spherical polar vector components of the far field beam
can be obtained from the paraxial scalar expressions for the
two plane polarised components $E_x$ and $E_y$:
\begin{eqnarray}
E_\theta & = & - E_x \cos\phi - E_y \sin\phi,\\
E_\phi & = & - E_x \sin\phi + E_y \cos\phi.
\end{eqnarray}

\subsection{TEM$_{00}$ beam}

The far field spherical wave limit of the paraxial Gaussian beam
formula~(\ref{paraxial_gaussian}) can be found by converting the cylindrical
coordinates to spherical coordinates and taking the large $r$ limit, giving
\begin{equation}
U = U_0 \exp( -k^2 w_0^2 \tan^2\theta / 4 )
\end{equation}

\begin{figure}[htb]
\centerline{\includegraphics[width=\columnwidth]{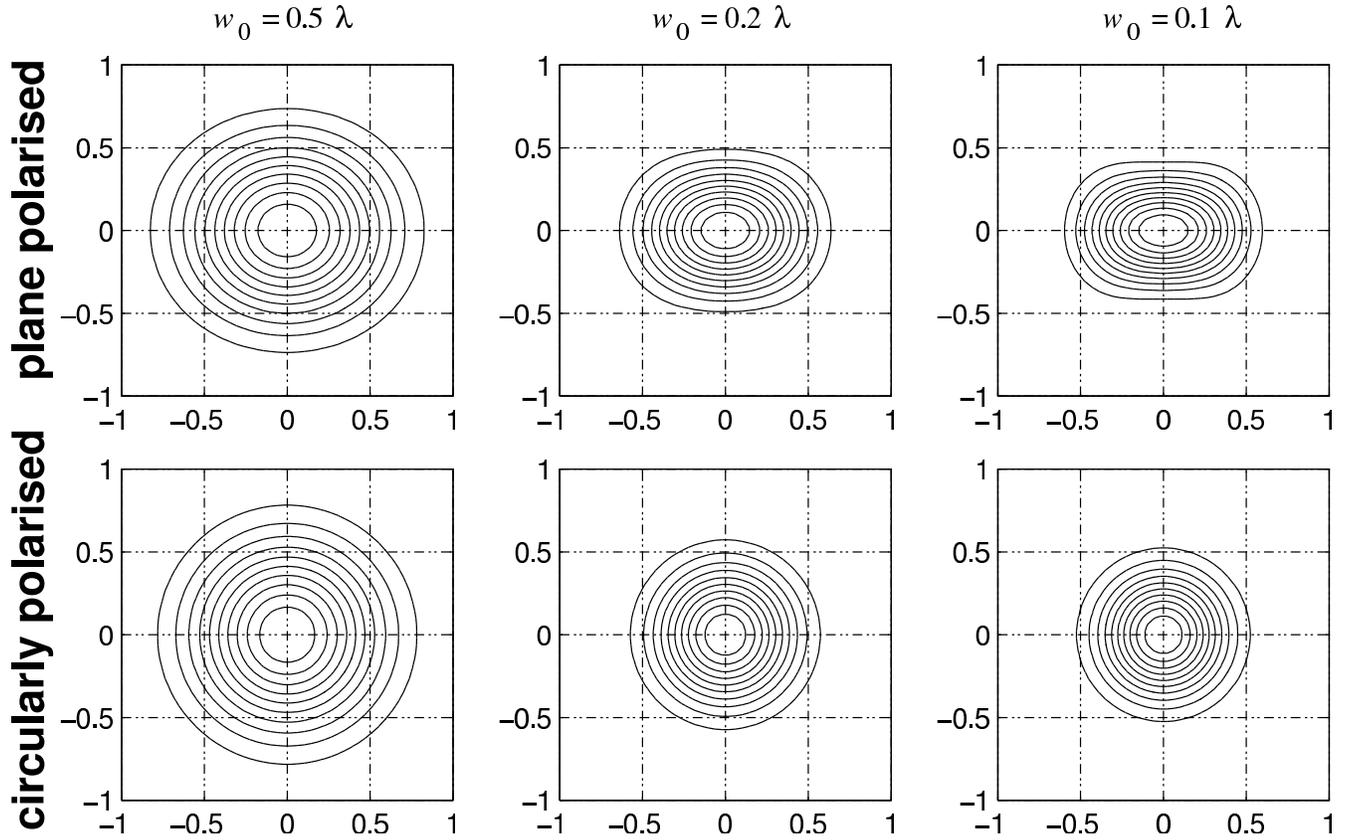}}
\caption{Irradiance contours in the focal plane of multipole beams far
field matched with TEM$_{00}$ beams. The top row shows plane-polarised
beams, with the plane of polarisation in the horizontal plane, and the bottom
row shows circularly-polarised beams. The beam waists vary from left to right as
$w_0 = 0.5\lambda$, $w_0 = 0.2\lambda$, and $w_0 = 0.1\lambda$. All distances
are in units of the wavelength.}
\label{far_6plot}
\end{figure}

Focal plane irradiance contours for far field matched beams are shown in
figure~\ref{far_6plot}. Except for the use of the far field for matching, the
procedure, and the results, are the same as the focal plane matching case.

\subsection{LG$_{pl}$ beams}

We recall that the focal plane matching procedure is greatly complicated when
the Poynting vector in the focal plane is not parallel to the beam axis, and
note that the same does not apply to far field matching---the Poynting vector
is always purely radial in the far field regardless of its behaviour in the
focal plane. Therefore, beams of these types, such as Laguerre-Gaussian
LG$_{pl}$ doughnut beams are best dealt with by far field matching. (The
failure of naive attempts to model LG$_{pl}$ beams by focal plane matching
simply by using the correct irradiance distribution with the correct
$\exp(\mathrm{i}m\phi)$ azimuthal phase variation for $E_x$ is readily observed
when attempted.) The far field limit for Laguerre-Gaussian beams~\cite{siegman}
is
\begin{equation}
U = U_0 (2\psi)^{(l/2)} L_p^l(2\psi) \exp( \psi + il\phi )
\end{equation}
where $\psi = -k^2 w_0^2 \tan^2\theta/4$ and $L_p$ is the generalised Laguerre
polynomial.

\begin{figure}[htb]
\centerline{\includegraphics[width=\columnwidth]{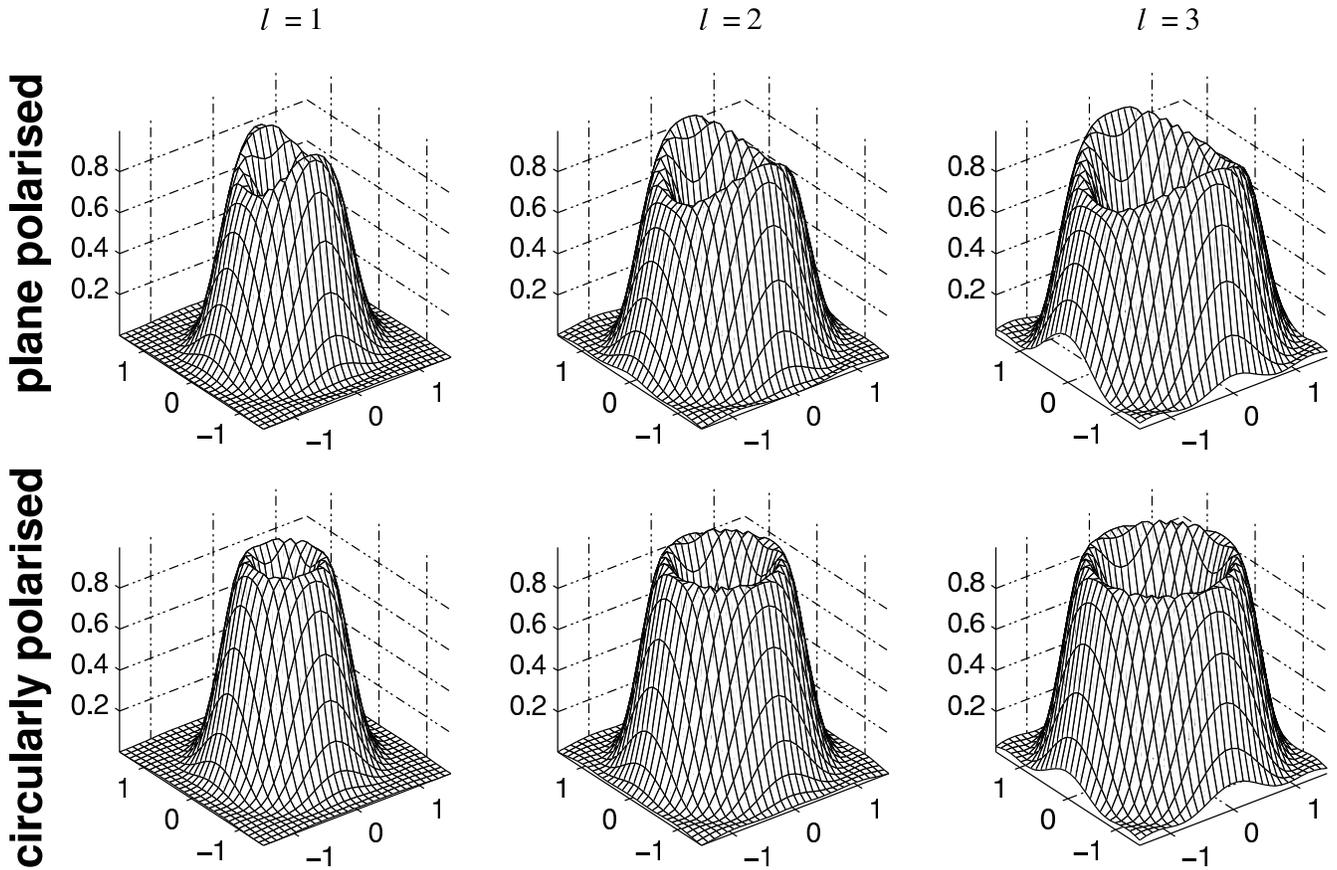}}
\caption{Focal plane irradiance of multipole beams far field matched with
LG$_{0l}$ beams. The top row shows plane-polarised beams, with the plane of
polarisation parallel to the lower right axis, and the
bottom row shows circularly-polarised beams. The beam waists is $w_0 =
0.5\lambda$ for all cases, and the azimuthal mode index $l$ varies from left to
right as $l=1$, $l=2$, and $l=3$. All distances are in units of the wavelength.}
\label{far_6lg}
\end{figure}

\subsection{Truncation by an aperture}

\begin{figure}[htb]
\centerline{\includegraphics[width=\columnwidth]{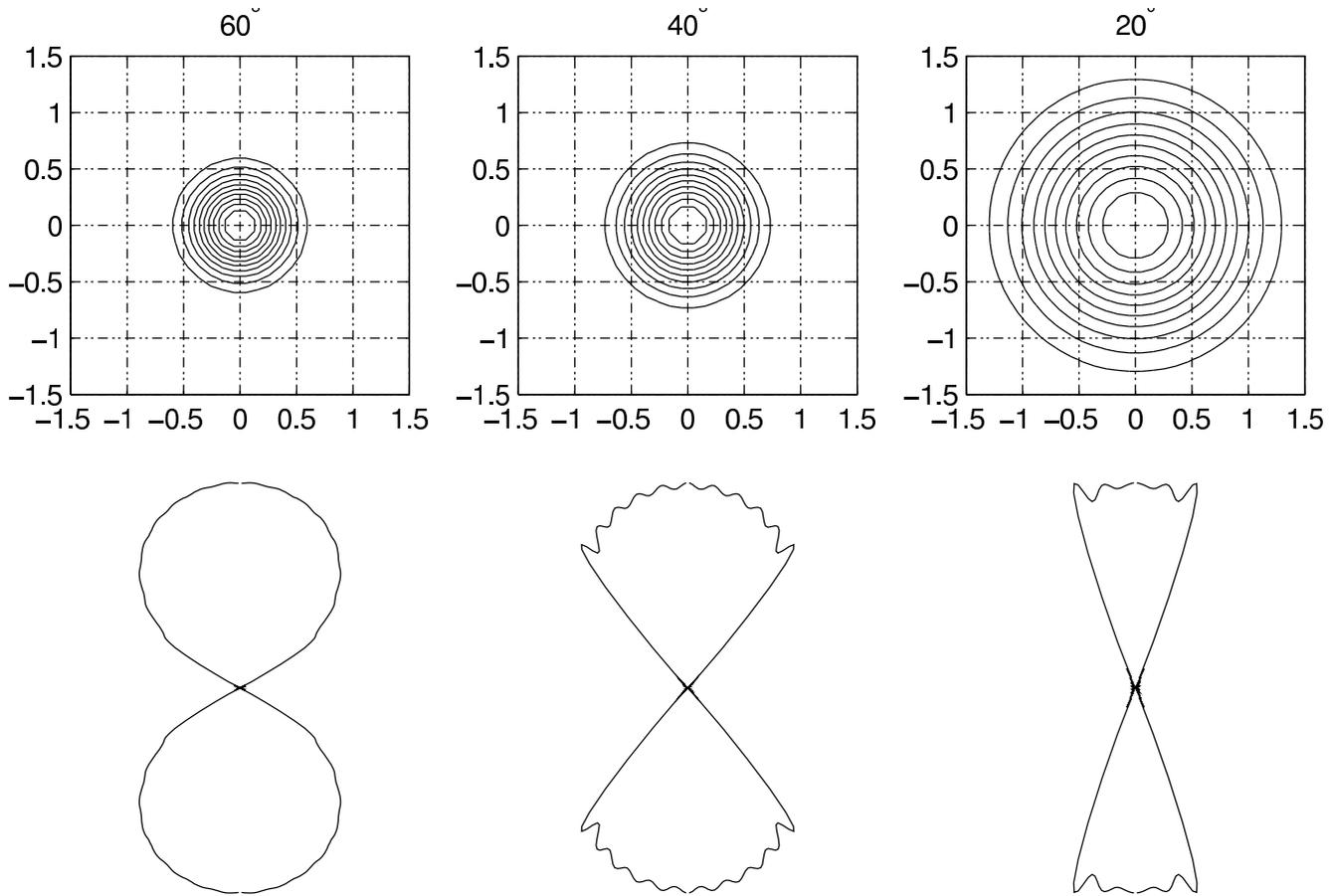}}
\caption{Circularly polarised TEM$_{00}$ beam of waist radius $w_0 =0.2\lambda$
truncated by hard-edged apertures restricting the incoming beam to within
$60^\circ$, $40^\circ$, and $20^\circ$ of the $-z$ axis,  shown from left to
right. The focal plane irradiance contours (top) and the radiation patterns
(bottom) are shown. All distances are in units of the wavelength.}
\label{far_trunc}
\end{figure}

Truncation by an aperture is readily included by restricting non-zero values of
the incoming beam to less than the angle of the aperture. If a hard-edged
aperture is assumed, a much higher $N_{\mathrm{max}}$ is required to accurately
truncate the field.
The aperture must sufficiently far away for the far field limit to be accurate,
which requires a distance $r \gg n^2/k$.
A soft-edged aperture may also more realistically model the
actual physical system. The effect of increasing truncation of a beam is shown in
figure~\ref{far_trunc}. The beam is a circularly polarised TEM$_{00}$ beam of
waist radius $w_0 =0.2\lambda$. Axisymmetry was assumed, with
$N_{\mathrm{max}} = 48$, with the solution of the linear system requiring
0.79\,s on a 1.5\,GHz PC. The radiation patterns show that the hard-edged
aperture is modelled with reasonable, but not perfect accuracy. The errors due
to using a finite number of VSWFs to model the sharp edge are exactly analogous
to those seen when using a finite number of Fourier terms to model a sharp step.
The increase in focal spot size due to diffraction is clearly shown.

\section{Conclusion}

The point-matching method can be successfully used to obtain multipole
expansion equivalents of focussed scalar paraxial beams. Multipole
expansions are required
for scattering calculations using the \textit{T}-matrix method or GLMT,
or calculations of optical forces and torques using these method, and to be
able to obtain a satisfactory expansion of a strongly focussed laser beam
that is equivalent in a meaningful way to a standard paraxial laser beam
is highly desirable.
The method is
fastest when the beams are strongly focussed, since the maximum degree
$N_{\mathrm{max}}$ required for convergence is smaller. The method appears
usable even when the beam is focussed to the maximum possible extent.
Truncation of the beam by apertures is readily taken into account. If it is
necessary to truncate the multipole expansion at $N_{\mathrm{max}}$ less than
the value ideally required for exact representation of the beam (for example, if
the \textit{T}-matrix of the scatterer is truncated at this $N_{\mathrm{max}}$),
the multipole expansion given by the point-matching method will be the best fit,
in a least squares sense, obtainable for this $N_{\mathrm{max}}$.
If $N_{\mathrm{max}}$ is sufficiently large, the multipole expansion will
be well-convergent, and the multipole field can be made to be arbitrarily close
to the far field/focal plane field with which it is matched. While the
multipole beam will not be equal to the original paraxial beam over all space
(only on the matching surface), the multipole beam can be considered as
a non-paraxial version of the standard beam.

While the emphasis in this paper has been on obtaining multipole expansions
of standard beams, the method used is applicable to arbitrary beams, 
including analytical forms of beams with corrections for non-paraxiality.
Since multipole expansions of such beams are still required for the
types of scattering calculations considered here, the point-matching method
may prove useful applied to such beams. The only restriction on the beam in
question is that it must be possible to calculate the fields at a suitable
representative set of points.

Compared with integral methods, point-matching is faster for strongly
focussed beams since the method tolerates a wider grid point spacing. The chief
disadvantage, compared with integral methods, is worse performance for
sufficiently large $N_{\mathrm{max}}$.

Compared with the localised
approximation~\cite{gouesbet1995,gouesbet1996b,polaert1998ao},
point-matching is slower, but is applicable to extremely focussed beams, and
allows simple calculation for arbitrary beams, including beams with no known
analytical representation.

Practical applications typically require multipole expansions of the
beam in a coordinate system centred on a scattering particle, rather than
the beam waist as done in the examples presented here. There are two
distinct methods in which particle-centred expansions can be determined.
Firstly, it should be noted that there is no requirement in the
point-matching method for the beam waist to coincide with the $xy$ plane
and the beam axis to coincide with the $z$ axis. Therefore, coordinate
axes can be chosen to coincide with the scattering particle, and the
points in the focal plane or far field at which the fields are matched
specified in this particle-centred coordinate system. This is simply done
for focal plane matching. For far field matching, we note that
translation of the coordinate system by $\mathbf{x}$ is equivalent to
a phase shift of $k \mathbf{x}\cdot\hat{\mathbf{r}}$ at the points in the far
field. A larger $N_{\mathrm{max}}$ will typically be required for
convergence since a larger radius is required to contain the beam waist
when the beam waist is not centred on the origin.

Alternately, and better for repeated calculations,
the rotation transformation for VSWFs
\begin{equation}
\mathbf{M}_{nm}^{(1,2)}(k\mathbf{r}) =
\sum_{m' = -n}^n D_{m'm}^n(\alpha\beta\gamma)
\mathbf{M}_{nm'}^{(1,2)}(k\mathbf{r}')
\end{equation}
and similarly for $\mathbf{N}_{nm}^{(1,2)}(k\mathbf{r})$ can be used, where
$D_{m'm}^n(\alpha\beta\gamma)$ are Wigner $D$
functions~\cite{mishchenko1991,varshalovich}, along with the translation
addition theorem~\cite{doicu1997ao,brock,varshalovich}.

\begin{figure}[htb]
\centerline{\includegraphics[width=\columnwidth]{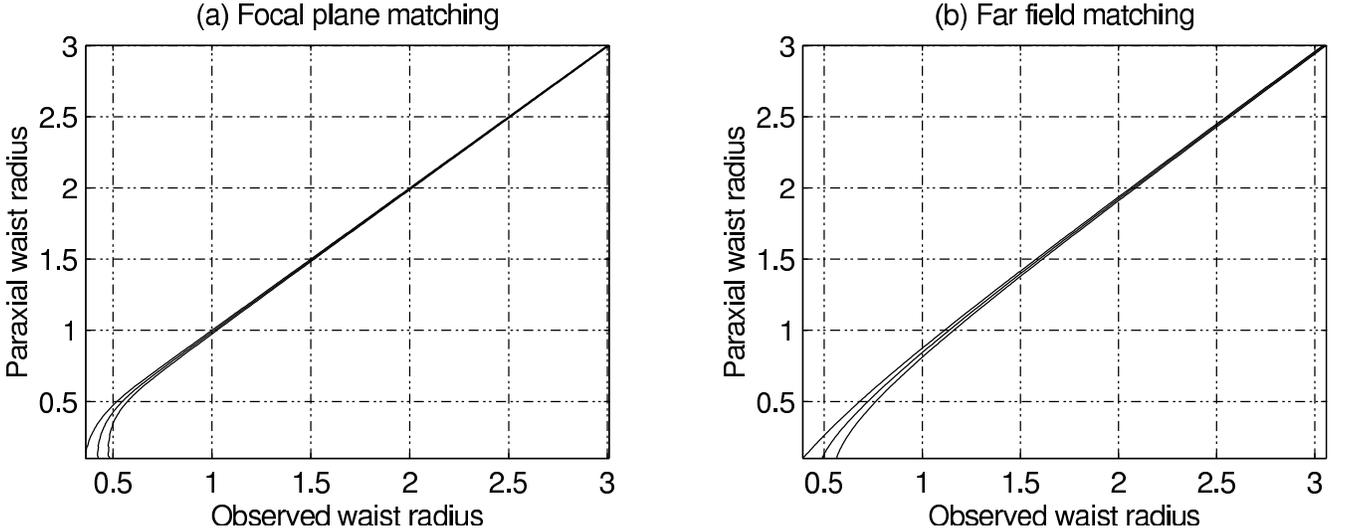}}
\caption{Comparison between paraxial and observed beam waists. Results are
shown for TEM$_{00}$ Gaussian beams matched in the focal plane (a), and
the far field (b). The three curves on each graph correspond to the beam waist
observed along the plane of polarisation (right), and perpendicular to the plane
of polarisation (left) for plane polarised beams, and the azimuthally
independent waist for circularly polarised beams (centre). Note that since the
electric field magnitude is Gaussian, the beam waist radius is the point at
$E(r) = E_0/\exp(-1)$, or, in terms of irradiance,  when $I(r) = I_0/\exp(-2)$.}
\label{focal_waist}
\end{figure}

We also note that the paraxial beam waist is a rather misleading parameter to
use the describe the multipole beam given by the point-matching procedure, since
the actual beam wasit of the multipole beam will differ from that of the
paraxial beam. To be able to caclulate a multipole beam of a given waist
radius from a paraxial beam, it is necessary to know what paraxial beam waist
corresponds to the actual beam waist. Graphs comparing the paraxial and observed
waists are given in figure~\ref{focal_waist}. For
computational convenience, approximation formulae of the form
\begin{equation}
w_{0\mathrm{paraxial}} = w_0 + \frac{c_1}{w_0}  + \frac{c_2}{w_0^2}
 + \frac{c_3}{w_0^3} + \cdots
\label{approximation_formulae}
\end{equation}
can be used to determine the required paraxial beam waist. The approximation
coefficients are listed in table~\ref{approximation_coefficients}.

\begin{table}
\begin{center}
\begin{tabular}{lr@{.}lr@{.}lr@{.}lr@{.}lr@{.}lr@{.}l}
\hline
& \multicolumn{2}{c}{$c_1$} & \multicolumn{2}{c}{$c_3$}
& \multicolumn{2}{c}{$c_3$} & \multicolumn{2}{c}{$c_4$}
& \multicolumn{2}{c}{$c_5$} & \multicolumn{2}{c}{$c_6$}\\
\hline
N$|$     & -0&01798   & -0&05457  & 0&1545    & -0&2102  & 0&1367  & -0&03405 \\
N$\perp$ & -0&0007615 & 0&004553  & -0&01072  & 0&01111  & -0&004148 \\
N$\circ$ & -0&01245   & -0&004407 & 0&01929   & -0&03468 & 0&02752 & -0&008022 \\
F$|$     & -0&1792    & 0&01347   & -0&04588  & 0&0393   & -0&0168 \\
F$\perp$ & -0&1265    & -0&001236 & 0&002310  \\
F$\circ$ & -0&1516    & -0&002584 & 0&0002883 & -0&002711 \\
\hline \\
\end{tabular}
\end{center}
\caption{Coefficients for approximation formulae~(\ref{approximation_formulae}).
The following symbols are used to describe the beam: N -- focal plane matched;
F -- far field matched;
$|$ -- plane polarised, along the direction of polarisation;
$\perp$ -- plane polarised, perpendicular to the direction of polarisation;
$\circ$ -- circularly polarised}
\label{approximation_coefficients}
\end{table}

In conclusion, a reasonable solution to the problem of multipole
representation of strongly focussed laser beams is presented. Although the
VSWF expansions are not identical to the paraxial beams from which
they are derived, they are related in a natural manner. These beams are of
particular interest for optical trapping and scattering by single particles
within optical traps.

\end{document}